\documentclass[useAMS,usenatbib]{mn2e}
\usepackage{graphicx}
\usepackage{epstopdf}
\usepackage{subfigure}
\usepackage[a4paper]{hyperref}
\usepackage{amssymb}
\usepackage{psfig}

 \def\be{\begin{equation}}
 \def\ee{\end{equation}}
\def\solmas{{{\rm M}$_\odot$}}
\def\simless{\mathbin{\lower 3pt\hbox
   {$\rlap{\raise 5pt\hbox{$\char'074$}}\mathchar"7218$}}}   % < or of order
\def\simgreat{\mathbin{\lower 3pt\hbox
   {$\rlap{\raise 5pt\hbox{$\char'076$}}\mathchar"7218$}}}   % > or of order
\def\etal{{\rm et al.}}

\def\solm{{\rm M}_\odot}

\title[Origin of star formation rates] {Shocks, cooling and the origin of star formation rates in spiral galaxies}
\author[I. A. Bonnell \etal]{Ian~A.~Bonnell$^1$\thanks{E-mail: iab1@st-and.ac.uk}, Clare L. Dobbs$^2$, and Rowan J. Smith$^3$\\
\\
\normalsize{$^{1}$ Scottish Universities Physics Alliance (SUPA),
School of Physics and Astronomy,}\\ 
\normalsize{University of  St. Andrews, North Haugh, St. Andrews, Fife KY16 9SS, UK}\\
\normalsize{$^{2}$ School of Physics, University of Exeter, Stocker Road, Exeter, EX4 4QL}\\
\normalsize{$^{3}$  Zentrum f\"ur Astronomie der Universit\"at Heidelberg, }\\
\normalsize{Institut f\"ur Theoretische Astrophysik, Albert-Ueberle-Str.\ 2, 69120 Heidelberg}\\
}

\date{\today}

\begin{document}

\date{Accepted 2012 December 28. Received 2012 November 15; in original form 2012 August 29}

\pagerange{\pageref{firstpage}--\pageref{lastpage}} \pubyear{2009}

\maketitle

\label{firstpage}

\begin{abstract}
Understanding star formation is problematic as it originates in the large scale dynamics of a galaxy but occurs
on the small  scale of an individual star forming event. This paper  presents the first numerical simulations to resolve the star formation process 
on sub-parsec scales, whilst also following the dynamics of the interstellar medium (ISM) on galactic scales.  
In these models, the warm low density ISM gas flows into the spiral arms where orbit crowding produces the
shock formation of dense clouds, held together temporarily by their external pressure. 
Cooling allows the gas to be compressed to sufficiently high densities that local regions collapse under their own gravity and form stars.
The star 
formation rates follow a Schmidt-Kennicutt $\Sigma_{\rm SFR}\propto \Sigma_{\rm gas}^{1.4}$ type relation with 
the local surface density of gas while following a linear relation with the cold and  dense gas.
Cooling is the primary driver of star formation and the star formation rates as it determines the amount of cold gas available for gravitational collapse. 
The star formation rates found in the simulations are offset to higher values relative to the extragalactic values, implying  a constant reduction, such as from feedback or magnetic fields, is likely to be required. 
 Intriguingly, it appears that a spiral or other convergent shock and the accompanying thermal instability can explain how star formation is triggered, generate the  physical
 conditions of molecular clouds and explain why star formation rates are tightly correlated to the gas properties of  galaxies.
\end{abstract}


\section{Introduction}

Star formation is the primary driver of galactic evolution and provides the birth place for planetary systems, yet it is poorly understood.
Observationally,  a strong correlation has long been found between the amount of gas in the galaxy and the rate at which stars form \citep{Schmidt1959,Kennicutt1989,KennicuttEvans2012}, whereby
the rate of star formation per unit area in the galactic disc is correlated with the (total) surface density of gas in the disc as $\Sigma_{\rm SFR} \propto \Sigma_{\rm gas}^{1.4}$. 
This (Schmidt-Kennicutt) relation holds both globally in galaxies over a large range of galaxy properties as well as locally on kpc scales of individual galaxies \citep{Kennicutt2007}.
In addition,  a near linear relation has been found between the dense molecular gas and the star formation rates \citep{Kennicutt2007,Bigiel2008,Wu2010,Schruba2011}, although it has recently been argued that systematic variations could exist between individual galaxies \cite{Shetty2012b}.
These correlations have inspired many theoretical ideas involving everything from 
gravitational instabilities, spiral shocks, local collapse, self-regulation through feedback 
\citep{Silk1997,Elmegreen2002,Li2006,Krumholz2009,Ostriker2011} but until now there has been no means of testing these various ideas
and thus constructing a complete theory of star formation in galaxies.

Recently, deviations from the above star formation relations have been found on smaller extragalactic ($\simless 200$ pc) scales  in the form of large scatter
at low gas surface densities \citep{Schruba2010,Onodera2010}. 
These deviations could be due to intrinsic variation in the cloud to cloud star forming properties, sampling issues
when rates are estimated from integrated fluxes due to massive stars, or  to the  
movement of newly formed stars out of their natal environments.   In contrast,
star formation rates in nearby Galactic molecular clouds, where the rates are calculated directly from the young stars present  and cloud masses from extinction \citep{Evansetal2009,Heiderman2010}, show instead a threshold at low surface densities before converging to a more standard Schmidt-Kennicutt relation (see also \citealt{Lada2012}). 

\begin{table*}
\centering
\begin{tabular}{c|c|c|c|c|c|clc}
 \hline 
Model & $\Sigma_{\rm disc}$ & Particle mass & Region size & Start time & Total time \\
 &  (M$_{\odot}$ pc$^{-2}$) & (M$_{\odot}$) & (pc) & (Myr) & (Myr) \\
 \hline
Global simulation \hfil  &4.2 & 40 & 20,000 & 0 & 370 \\
Cloud re-simulation \hfil&4.2 & 0.156 & 2000 & 310 & 54 \\ 
Gravity re-simulation \hfil&4.2 & 0.156  & 250 & 350 & 5\\ 
\hline
Cloud re-sim high\hfil&42.4 & 1.56 & 2000 & 345 & 17 \\
Gravity re-sim high\hfil&42.4 & 1.56  & 250 & 350 & 2\\ 
\hline
Cloud re-sim low \hfil&0.42 & 0.0156 & 2000 & 345 & 10 \\
Gravity re-sim low \hfil&0.42 & 0.0156  & 250 & 350 & 8\\
\hline
\end{tabular}
\caption{This table displays the details of the simulations presented. The times shown are in relation to the global simulation. 
The Gravity and low and high re-simulations are all derived from the Cloud re-simulation.  The total masses  are 2.5$\times 10^9$ M$_{\odot}$ 
for the global disc simulation, 1.71$\times 10^6$ for the Cloud re-simulation and 1.61$\times 10^6$
for the Gravity re-simulation. For the low and high re-simulations the total masses are ten times lower ($\approx 10^5\solm$)  and higher ($\approx 10^7\solm$)
than  for the Cloud and Gravity re-simulations, respectively.}
\label{runs}
\end{table*}

Star formation in galaxies involves the collection of gas into dense molecular clouds, and the collapse of localised regions of these clouds to form stars, and radiative and kinetic feedback from the
young stars back into the interstellar medium, processes which may or may not be simultaneous. Galactic scale numerical simulations have shown that in spiral galaxies, molecular clouds likely form through the coalescence of less dense clouds in spiral shocks \citep{Dobbs2008}, aided by self gravity for the most massive clouds. 
Several studies have attempted to study the Schmidt-Kennicutt relation driven by combinations of self-gravity and feedback \citep{Li2006,Wada2007,Tasker2006,Dobbs2009,Koyama2009,Dobbs2011,Kim2011,Agertz2011,Shetty2012},
but were unable to resolve the scale of star formation and required recipes for star formation and sub-grid modelling to account for the combined effects of cooling and stellar feedback.

On the smaller, parsec scales of individual molecular clouds, local simulations  do resolve star formation \citep{SmiLonBon2009,Bonnell2011,Krumholz2010,Bate2012}, but these start with highly idealised conditions that do not reflect the physics of the cloud formation process.  Feedback has also been included on local scales, but does not appear to significantly alter the star formation rates of individual molecular clouds \citep{Dale2008,Dale2011,Dale2012}. Cloud-cloud collisions \citep{Whitworth1994,Kitsionas2007} and colliding flows simulations provide a mechanism for triggering star formation 
\citep{Koyama2000,Heitsch2006,Vazquez2007,Banerjee2009,Clark2012} but lack the galactic context that could generate such flows \citep{Bonnell2006}.

This paper presents the first numerical simulations capable of following both the large-scale galactic
flows of the gas and still resolve down to the small scale physics of star formation. 
We use a series of simulations to probe from the large scale of a galaxy to the small scale of a star forming molecular cloud, in order to investigate what physics may lie behind  the observed Schmidt-Kennicutt relation. In this paper, we concentrate on the physics of cloud formation and self-gravity
as the possible mechanism behind the galactic star formation rates.

\section{Calculations}

\begin{figure*}
%\centerline{\includegraphics[scale=0.7,bb=0 140 600 700]{figure1.eps}}
%\centerline{\includegraphics[scale=0.65]{figure1ian.eps}}
\centerline{\includegraphics[scale=0.4]{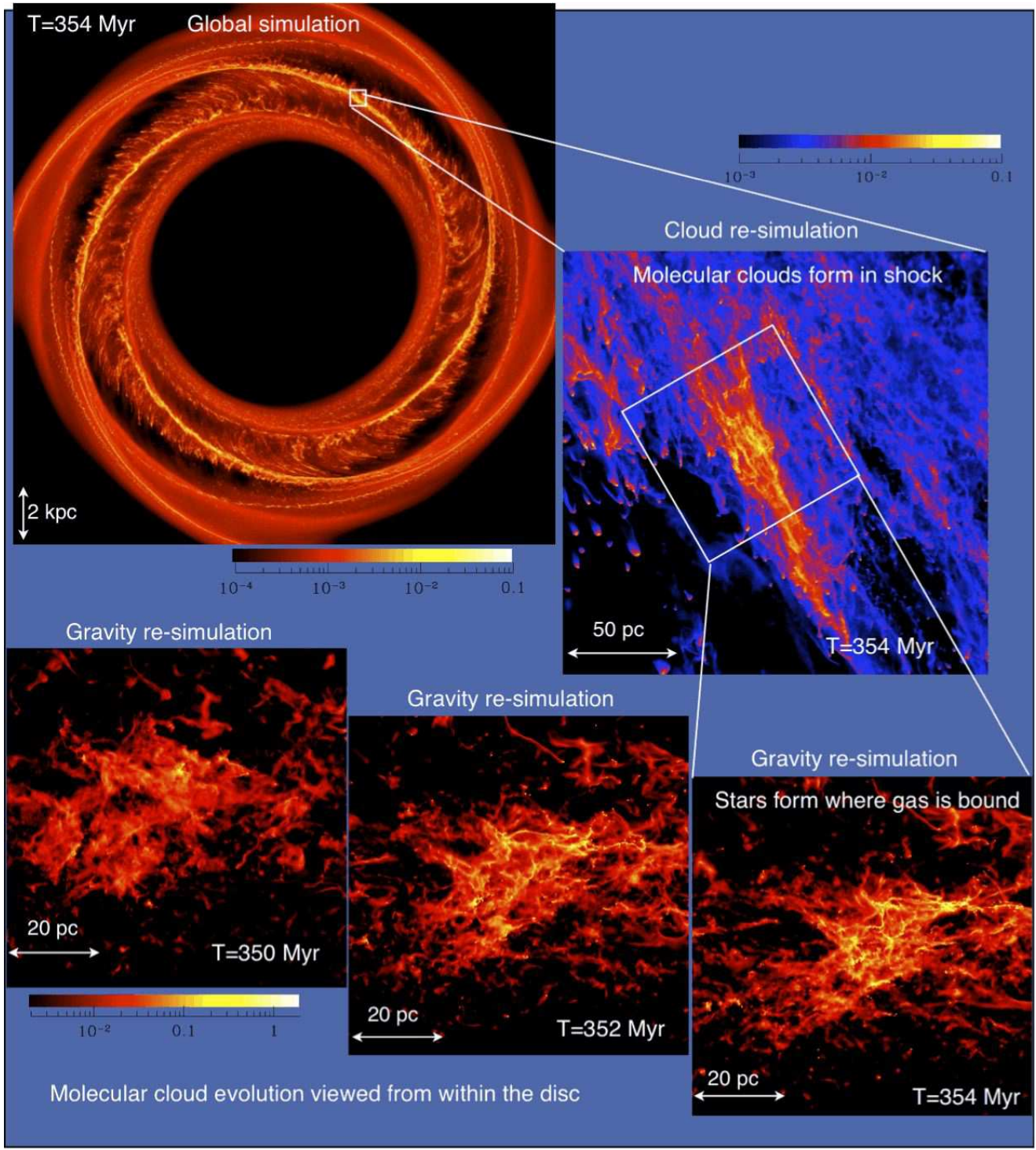}}
\caption{ An overview  of the numerical simulations with $\Sigma_{\rm disc} =4.2\solm$ pc$^{-2}$ is shown in terms of their column densities. In the global galactic disc simulation, gas shocks,
cools and coalesce in the spiral arms to form dense clouds. One of the clouds from the global simulation is used to perform the Cloud re-simulation, where we have increased the mass resolution by a factor of 256 in order to study in detail how the star formation region is formed. The last 3 panels show the 
cold gas ($T<50$ K) in the central regions of the Gravity re-simulation, which now includes self gravity and sink particles to indicate stellar clusters. Gas in the two re-simulations explores scales, and exhibits the densities and temperatures typical of molecular clouds ($10^3$ cm$^{-3}$ and $<20$ K) which are otherwise unfeasible in full global disc simulations. In each panel, the column density of the gas is shown, the logarithmic scale from $10^{-4}$ to  $0.1$ g cm$^{-2}$ for the global simulation, $10^{-3}$ to  $0.1$  g cm$^{-2}$ for the Cloud re-simulation, and from $0.002$ to $2$ g cm$^{-2}$ for the Gravity re-simulation.}
\end{figure*}

%\begin{figure*}
%\centerline{\includegraphics[scale=0.7,bb=0 140 600 700]{figure1.eps}}
%\centerline{\includegraphics[scale=0.5]{figure1ian.eps}}
%\centerline{\includegraphics[scale=0.6, bb=250 50 600 600]{figure1apjl.ps}}
%\centerline{\includegraphics[scale=0.6]{figure1apjl_cld.eps}}
%\centerline{\includegraphics[scale=0.6]{figure1apjl.eps}}
% \caption{ An overview  of the numerical simulations with $\Sigma_{\rm disc} =4.2\solm$ pc$^{-2}$ is shown in terms of their column densities. In the global galactic disc simulation, gas shocks,
% cools and coalesce in the spiral arms to form dense clouds. One of the clouds from the global simulation is used to perform the {\sl Cloud} re-simulation, where we have increased the mass resolution by a 
% factor of 256 in order to study in detail how the star formation region is formed. The last 2 panels show the 
% cold gas ($T<50$ K) in the central regions of the {\sl Gravity} re-simulation, which now includes self gravity and sink particles to indicate stellar clusters. Gas in the two re-simulations explores scales, and 
% exhibits the densities and temperatures typical of molecular clouds ($10^3$ cm$^{-3}$ and $<20$ K) which are otherwise unfeasible in full global disc simulations. In each panel, the column density of the gas 
% is shown, the logarithmic scale from $10^{-4}$ to  $0.1$ g cm$^{-2}$ for the global simulation, $10^{-3}$ to  $0.1$  g cm$^{-2}$ for the {\sl Cloud} re-simulation, and from $0.002$ to $2$ g cm$^{-2}$ for the {\sl 
% Gravity} re-simulation.}
%\end{figure*}

The numerical simulations used the  three-dimensional smoothed particle hydrodynamics (SPH) method to follow the gas dynamics in a galactic potential including 4 spiral arms \citep{DBP2006}. 
The simulations are used to  probe three, successively smaller scales, from the scale of a spiral galaxy, to the formation of dense clouds in the ISM, to the relatively small, sub-pc scale where star formation occurs.  
In addition to following the full galaxy disc simulation  for 360 Myrs, we carried out high-resolution simulations of a sub-region of the disc with and without self-gravity.
Our  high-resolution non self-gravitating Cloud re-simulations followed this region for the last 53.5 Myr  of the full disc evolution while our Gravity re-simulations including self-gravity followed the same region over 5 Myrs near maximum compression to determine the star formation rates. Self-gravity is only included in the last set of simulations due to the short timestep required to resolve high-density self-gravitating gas. Each smaller scale simulation uses the conditions from the larger scale to ensure that the global dynamics are included.  We neglect feedback from young stars and magnetic fields in these calculations in order to probe the 
effects of the spiral shock, cooling and gravity in triggering star formation and generating the star formation rates.
An overview of our fiducial simulations, with a mean surface density of $4.2$ \solmas\ pc$^{-2}$, is shown in Figure 1. In addition, we also repeated the Cloud, and then Gravity re-simulations at 10 times higher and lower surface densities (discussed in \S 6). All the calculations are detailed in Table~1 and discussed further in  Section 2.2.

\subsection{Thermal Physics}

\begin{figure}
\begin{center}
\includegraphics[width=8cm]{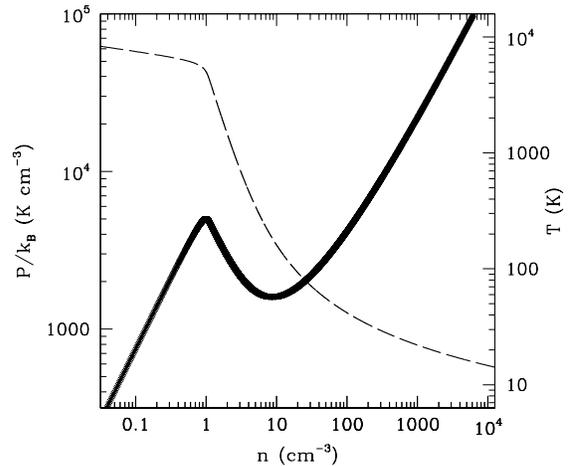}
\caption{\label{pressequil}The equilibrium temperature (broken line) and pressure is plotted as a function of gas density. The gas is nearly isothermal at $\approx 10^4$K  for low densities,
ensuring a high external pressure. Gas above densities of 1 cm$^{-3}$ cools and is compressed by the external pressure  until reaching densities of greater than 100 cm$^{-3}$ where pressure balance is restored. Cool gas at $T<1000$ K can be retained between shocks/collisions as long as the external pressure does not drop below $P/k_B \approx 1500$ K cm$^{-3}$.}
\end{center}
\end{figure}

The  thermodynamics of the gas includes compressional and shock heating as well as heating from a background interstellar radiation field. Cooling is included  through a cooling function
appropriate for interstellar gas that is optically thin, including atomic and molecular line cooling and cooling from dust
\citep{Koyama2002}. 
This cooling function is used to
calculate an equilibrium temperature and a cooling time to reach this equilibrium. Together this allows the internal energy to be integrated implicitly and
without affecting the hydrodynamical timesteps (for more details, see \citealt{Vazquez2007}). At low densities, 
$n < 1$ cm$^{-3}$,  the gas is nearly isothermal at $T\approx 10^4$ K. At higher densities, a cooling instability 
 occurs when atomic line cooling dominates at densities of $n>1 $cm$^{-3}$ while molecular and dust cooling is more important at the high gas densities appropriate for molecular clouds \citep{Koyama2002}. 
The cooling curve has equilibrium temperatures that produce a bi-stable ISM where a warm ($T\approx 10^4$K)
low density medium exists in pressure equilibrium with the cold ($T\approx 10$K) dense gas where star formation can occur 
\citep{Wolfire1995,Field1969}. This parametrisation neglects the detailed chemistry
and locally variable extinction and will likely overestimate the cooling rates at low densities (and overestimating the generation of self-gravitating gas and the star formation rates) while underestimating them  at higher densities  \citep{Clark2012, GloverClark2012b}.

The critical aspect of the simulations is the thermal instability that occurs when the gas  is compressed to densities $> 1$ particle cm$^{-3}$, and is best understood in terms of the gas pressure. Figure~\ref{pressequil}  shows the equilibrium pressure density relation for this thermal cooling law. The gas enters the shock at low densities and is compressed nearly isothermally at $\approx 10^4$ K, increasing its internal gas pressure. When gas is compressed beyond a critical density of 1 cm$^{-3}$, cooling dominates and the gas cools and is under-pressured until reaching densities above 100 cm$^{-3}$, the densities of star forming molecular clouds. The surrounding gas ensures a high external pressure and along with
the ram pressure of the shock, compresses the cooling gas to densities where self-gravity can take over.

Once the gas leaves a shock, the external pressure is reduced and the gas can return to the warm phase. If a
sufficiently high pressure is maintained, then some of the gas can be retained in a cool phase at densities near $n\approx 10$ cm$^{-3}$ and temperatures of $100<T<1000$K. This is likely to occur between collisions while the gas is still in a given spiral arm.  Subsequent collisions involving the cool gas will be stronger due to the now lower sound speed in the gas.

\subsection{Numerical Simulations}

\begin{figure*}
\begin{center}
\includegraphics[width=15cm]{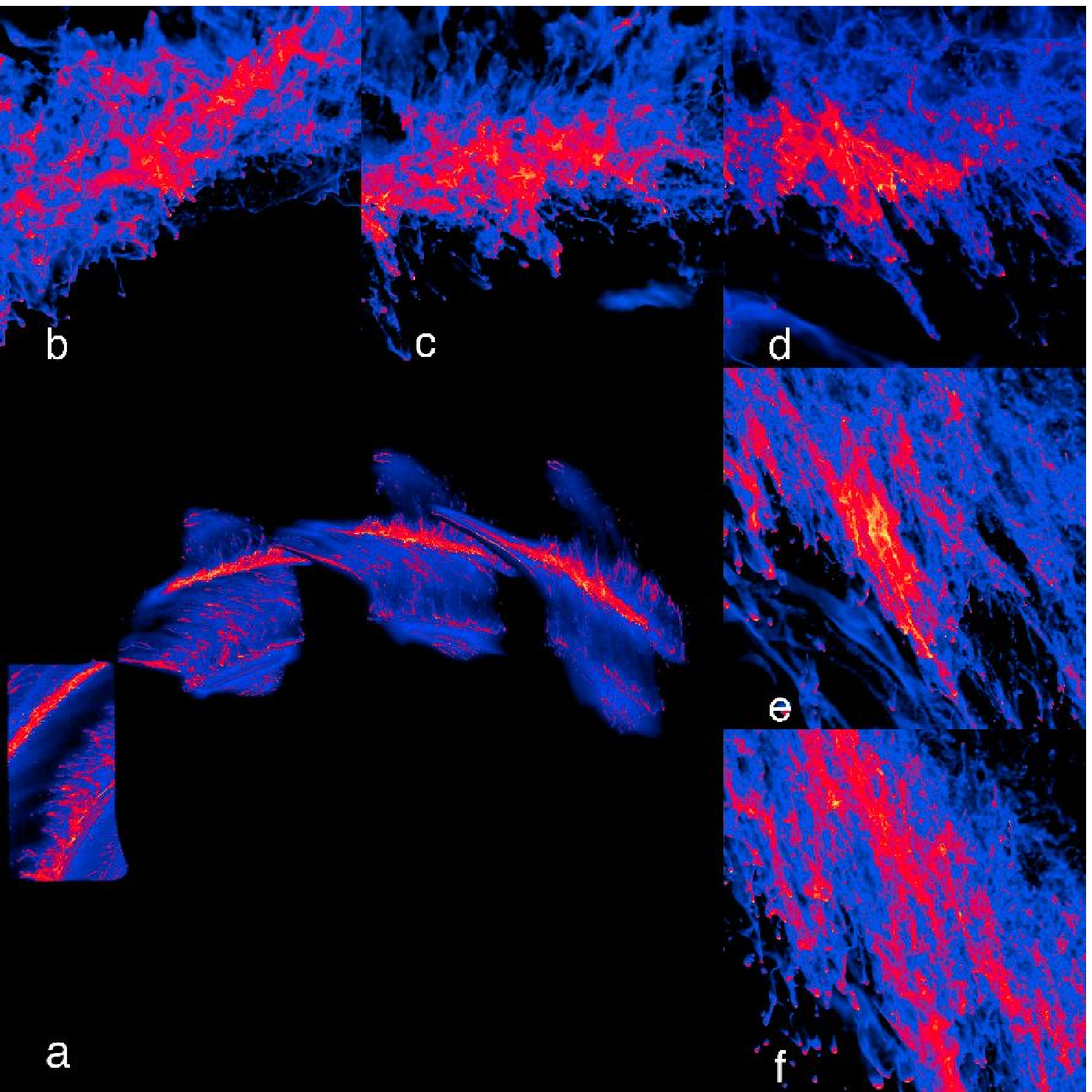}
\includegraphics[scale=0.33]{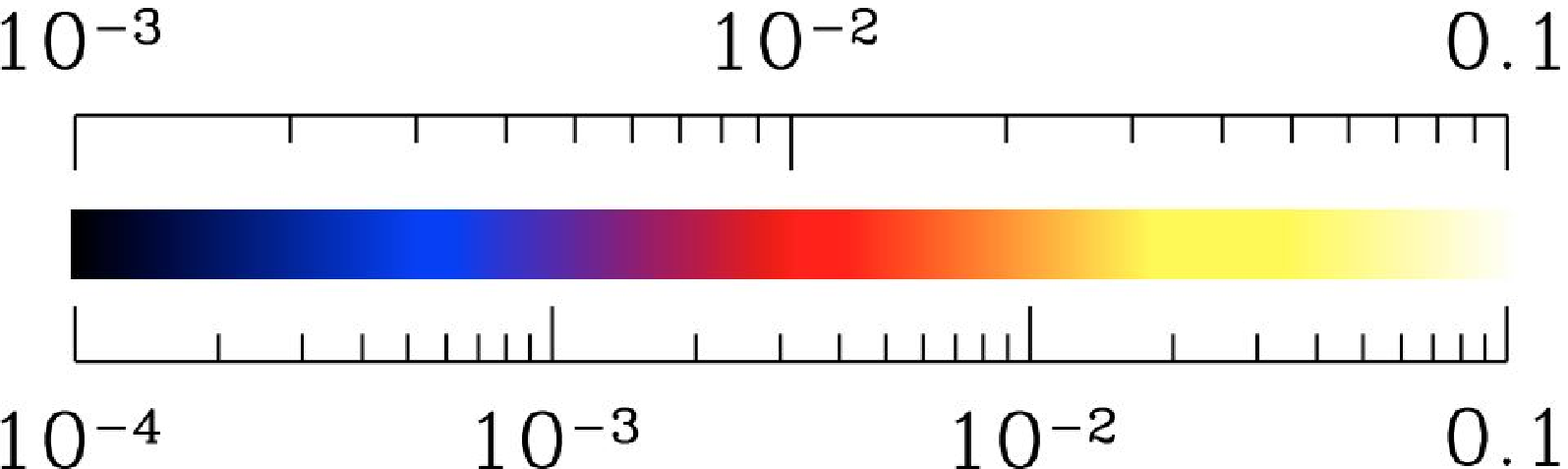}
\caption{\label{prettypics2}The  evolution of the high-resolution, non self-gravitating {\sl Cloud} simulation is shown  in panel a at 4 different times ( 307.7, 325.6, 333.1 and 353.8 Myr), 
as the gas rotates clockwise around the galaxy. The column densities are plotted logarithmically  from $10^{-4}$ to $0.1$ g cm${-2}$ (bottom scale) and the panel shows a region 12 kpc on a side.
The smaller panels (b-f) show the zooms of the evolution, from top left to bottom right, 
at times b) $17.9$ Myr, c) $25.4$ Myr, d) $33.0$ Myr, d) $44.1$ Myr and e) $53.5$ Myr from the start of the high-resolution simulation ($307.7$ Myr).
Each panel shows a 200 by 200 pc region centred on the  clump as it forms though the spiral shock and cloud coagulation.
The gas
is compressed into individual clouds which interact with each other and the inflowing cold gas. After maximum compression, the cloud is torn apart 
by the divergent flow as it leaves the spiral arm. The column densities are  shown logarithmically between a maximum of $0.1$ and a minimum of $0.001$ g cm$^{-2}$ (top scale).}
\label{fig:prettypics}
\end{center}
\end{figure*}

The global disc simulation used $2.5 \times 10^7$ SPH particles to model $10^9$ \solmas\ of gas in an annulus 5 to 10 kpc,
for an initial surface density of $\Sigma_{\rm disc} \approx 4.2$ \solmas\ pc$^{-2}$.
The gas was given an initial temperature of $10^4$ K, and allowed to cool during the evolution due to the thermal physics,
resulting in a multiphase ISM. 
The evolution is followed for over 350 Myrs to provide the initial conditions and boundary regions for the high resolution simulations.

The high resolution re-simulations were constructed by taking a high-density region from the global disc calculation, at a time of  350 Myr ($3.5 \times 10^8$ years),  and re-resolving this by splitting the original particles into 256 lower-mass particles. A 250 pc region of interest was chosen by eye, containing $1.71 \times 10^6$ \solmas\ of cold gas,
and then modeled using 1.09 $\times 10^7$ SPH particles with an individual particle mass of $0.156$ \solmas. 
The region was traced backwards through $\approx 1/4$ of 
a galactic orbit ($\approx 5 \times 10^7$ years) to produce the initial conditions for this high resolution  re-simulation. The region was augmented to include boundary particles, regular SPH particles  but outwith our region of interest, to ensure the hydrodynamics of our re-simulation are accurately modelled. These boundary particles include any particle that comes within 40 \% of the size of the region. Any of these boundary particles that come within 15 \% of the size of the region are split into 4 lower-mass particles, while those that come within 10 SPH kernel lengths of a particle in our region of interest are split into 16 lower-mass particles. Furthermore, if these particles come within 4 SPH kernel lengths (defined from the original simulation), they are split into 64 lower-mass particles. This ensures that the particles in our region of interest, and all other particles in the simulation,  only encounter other particles that are at most a factor of 4 different in mass.  The boundary particles are evolved as are the standard SPH particles. The rest of the galactic disc was not modeled in these Cloud, and Gravity re-simulations.
The Cloud re-simulation (region of interest plus boundaries) contained and $2.8 \times 10^7$ SPH particles and $3.09 \times 10^7$ \solmas. The simulation was then evolved forwards for 54 Myr  as the gas flowed back through the spiral shock to reproduce the region of interest at high resolution.

The Gravity re-simulations started using the inner 250 pc of the Cloud re-simulations at a time of $350.8$ Myr. 
Self-gravity is calculated via a tree-code \citep{Benzetal1990} and includes both the galactic potential and the gas contributions. 
The particle mass of $0.156$ \solmas\ in the Gravity re-simulation is adequate to resolve gravitational fragmentation
down to masses of $\approx 11$  \solmas. 
This is not sufficient to resolve the formation of individual low mass stars but is certainly sufficient to resolve the 
formation of stellar clusters where most stars form.   
These values are ten times lower and higher in the low and high re-simulations, respectively.

Star formation is modelled by sink particles \citep{Batesph1995}
where bound, collapsing regions are replaced by a single particle that can continue to accrete infalling gas that falls within $0.25$ pc.
Replacing a full SPH kernel containing some ~70 SPH particles ensures that the star formation is resolved and driven by the local gravitational collapse.
This allows us to derive the star formation rates rather than input a parametrised rate based on local properties as is commonly used  in cosmological simulations.
We assume a 100 \% efficiency of star formation for the gas that falls within our sink particles.  
Star formation efficiencies within $0.25$pc  could be as low as  10-50 \%. 
and hence we will determine an upper limit on star formation rates by a factor of 2-10.

For the two additional Cloud and Gravity re-simulations,  the mean surface densities in the disc were increased and decreased by an order of magnitude ($\Sigma_{\rm disc} = 0.42$ and $42$ \solmas\ pc$^{-2}$) compared to the standard $\Sigma_{\rm disc} = 4.2$ \solmas\ pc$^{-2}$ run. 
The global dynamics have minimal dependence on the different gas densities in the absence of self-gravity. However  the gas is slightly warmer/cooler  in the low/high
surface density re-simulations due to the dependence of heating and cooling on density. We thus reran the non self-gravitating Cloud re-simulations
for a minimum of 10 Myrs to ensure the gas temperatures were correct.
The Gravity re-simulations  were evolved for shorter and longer time periods for the higher and lower surface densities, respectively, reflecting
the different free-fall timescales of the self-gravitating gas involved. In all cases, the star formation rates had reached a steady level before
the end of the simulation.

\section{Results}
\subsection{{\sl Global} disc simulation}

The global disc simulation follows the gas, with an initial surface density of $\Sigma_{\rm disc}\approx 4.2$ \solmas\ pc$^{-2}$,  through 350 Myr  in which time it undergoes several (3-4) spiral arm passages. 
During the simulation, the gas is subject to the spiral potential, heating and cooling, but not self-gravity of the gas. The spiral arm potential produces convergent gas streams that meet and shock in the spiral arms \citep{DBP2006}. Each time the gas enters the spiral arms, the gas is compressed, cools and then re-expands in the inter-arm region. 
The cooling instabilities in the spiral arms lead to clumpy molecular cloud structures, which as the gas leaves the arm, are sheared into spurs or feathers
In addition, there are repeated collisions and internal shocks inside the spiral arms. This ensures that the structures we are studying are formed self-consistently,  and thus provide an appropriate starting point for studying how star formation is initiated in galactic discs.

 At the end of the global disc simulation (Fig.~1), $\approx 15$ \% of the gas mass ($1.5 \times 10^8$ \solmas) is sufficiently cold ($T<100$K) and dense ($n>100$ cm$^{-3}$) to provide the necessary environment for star formation. The thermal instability and the repeated compression generates a multiphase interstellar medium where cold, dense gas is in pressure equilibrium with the warm, low density gas \citep{Field1969}.

%The first simulation follows the full galactic disc modeled with $2.5 \times 10^7$ SPH particles to follow $10^9$ \solmas\ of gas in an annulus
%that extends from 5 to 10 kpc for a mean surface density of $\approx 4 \solm\ {\rm pc}^{-2}$.
%The gas is followed over $3.5 \times 10^8$ yrs as it orbits around the galaxy under the influence of the galaxy's gravitational potential including that of the spiral arms. This potential produces convergent gas streams that meet and shock in the spiral arms. The initially warm ($10^4$ K) gas goes  through several spiral arm passages  where it is compressed  into dense cold structures before  re-expanding in the inter-arm region (see also Dobbs \etal 2010). This ensures that the structures we are studying are formed self-consistently,  and thus provide an appropriate starting point for studying how star formation is initiated in galactic discs. Figure~\ref{prettypics} shows the final state of the global disc calculation where $\approx 15$ \% of the gas mass ($1.5 \times 10^8$\solmas) is sufficiently cold ($T<100$K) and dense ($n>100$ cm$^{-3}$) to provide the necessary environment for star formation.

\subsection{{\sl Cloud} re-simulation: Formation of Molecular  Clouds}

%A high resolution re-simulation of a dense cloud formed in the full disc calculation was used to establish the physical conditions for start formation. 

\begin{figure}
\begin{center}
\includegraphics[scale=0.45]{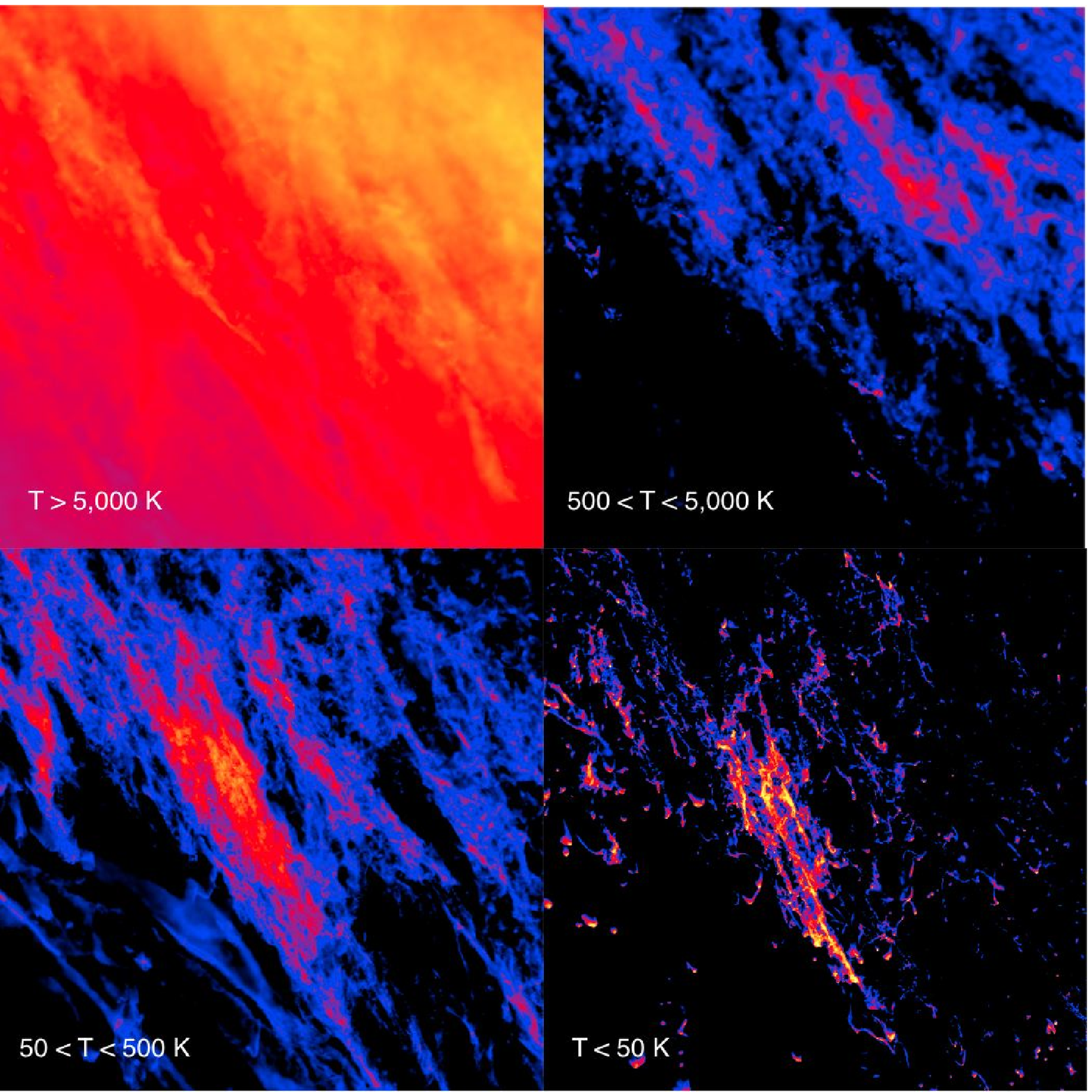}
\includegraphics[scale=0.33]{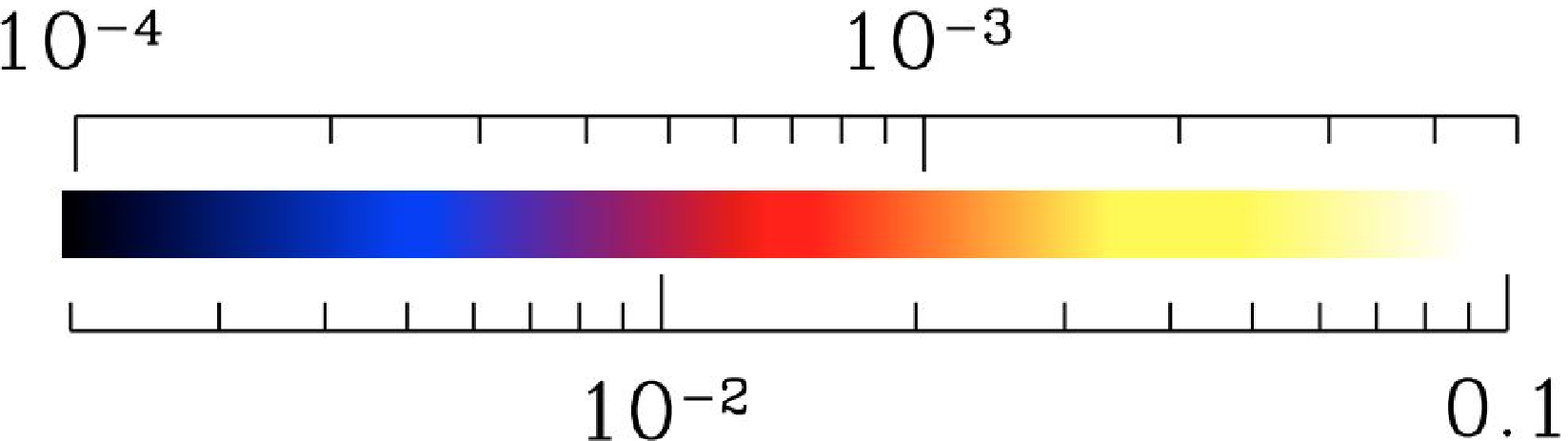}
\caption{\label{tempmaps}Column densities of the different gas as a function of temperature are shown at maximum compression in the non self-gravitating Cloud re-simulation. The hot ($T \ge 5000$K) and warm ($500\le T\le 5000$ K) 
gas is plotted between minimum and maximum column densities of $1\times 10^{-4}$ and $5 \times 10^{-3}$ g cm$^{-3}$ (top scale). The cool
($50\le T \le 500$K) and cold ($T\le 50$K) gas is plotted  between minimum and maximum column densities of $2\times 10^{-3}$ and $1 \times 10^{-1}$ g cm$^{-3}$ (bottom scale). }
\label{fig:temp}
\end{center}
\end{figure}

Figure~\ref{prettypics2} shows our standard ($\Sigma\approx 4.2$ \solmas\ pc$^{-2}$) high resolution Cloud re-simulation as the gas evolves around the galactic disc over 54 Myr, along with several zoomed snapshots of the gas in its centre of mass frame.
This shows how the gas clouds form through infall in the spiral shock and cloud coagulation in the spiral arm. 
The cloud complex consists of material which has been in the spiral arm for over 10 Myr
and gas which has entered from the inter-arm region. Gas from the spiral arm is already cold and dense, whilst gas entering comprises warm ($ > 8000$K) gas and some cooler and denser  gas from previous spiral arm encounters. These large scale flows of gas converge to produce a complex structure of dense gas clouds including a central ($\sim 40 \times 80$ pc) cloud (see Figure~1). The clouds are not  well defined objects as material is continually entering and leaving them.
The cloud is eventually shredded by the galactic tidal field after 53.6 Myrs. 

At each shock or collision, the gas is compressed and cools while maintaining pressure equilibrium with the surrounding warm gas. The cool gas is embedded in the warm, low density phase (Figure~\ref{tempmaps}) and subsequent shocks involve this clumpy medium. If the external pressure decreases (i.e. after a shock), the cool gas can re-expand. If the pressure drops below the critical level to maintain a bistable gas, all the cool gas returns to the warm phase. If the external pressure is instead maintained above this critical level, then some of the gas will remain cool. This is especially true for the multiple collisions inside a spiral arm.  

%Some of  the gas is already in  the spiral arm at the start of the simulation. 
%The evolution is driven by a combination of gas infalling into the spiral arm and by the compression and merger of other clouds in the vicinity. 
%The incoming gas is comprised of warm ($ > 8000$K) gas which cools as it is compressed in the spiral shock, mixed with
%some  cooler and denser  gas from previous spiral arm encounters. %Ten percent of the gas, including boundary regions, is  compressed sufficiently to become cold ($T<100$K) and dense ($n > 100$ particles cm$^{-3}$) to form molecular clouds.
%The dense clouds are formed from the compression of predominantly warm  ($ > 8000$K), low density gas.  The shocked gas cools rapidly and collapses at constant external pressure until reaching temperatures 
%$ of $\le 50$K. 

In this {\sl Cloud} re-simulation, approximately half of the mass (excluding boundary particles) reaches molecular cloud densities and temperatures. The high density regions are  surrounded by a  cocoon of warm gas 
which provides the mass reservoir from which they form, but 
crucially also helps hold them together. 
The cold dense gas occupies a hole in the warm gas  (Figure~\ref{tempmaps}).

The high velocity
shock and cooling also produce a turbulent and highly structured environment 
(see also \citealt{Bonnell2006,Heitsch2006,Dobbs2007}). 
The turbulence provides an important global support to the clouds, while the structure provides the seeds for local gravitational collapse.
The dense gas leaves the spiral arm some 6 Myr after maximum compression and is torn apart by the galactic shear, setting an upper limit on the lifetimes of molecular clouds \citep{DBP2006}.

\subsection{{\sl Gravity} re-simulation: Following Star Formation}

\begin{figure}
\begin{center}
\includegraphics[scale=0.4]{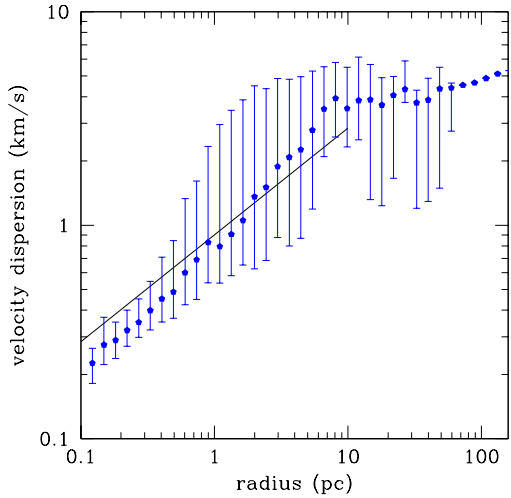}
\caption{\label{veldisp}The relation between the mean 1-D velocity dispersion and the sizescale of subregions in the cloud centred on the dense gas is
shown from the early stages of the Gravity re-simulation. {The rms scatter in the data is indicated by the errorbars. Points without errorbars represent  only single global values are possible at that size. The relation follows the typical Larson relation
for turbulent giant molecular clouds from 0.1 to 10 pc, but with significant scatter.} The Larson relation  $\delta v= 0.9 R_{\rm pc}^{0.5}$ (km/s) is also plotted as the solid line.}

\end{center}
\end{figure}

\begin{figure}
\centerline{\includegraphics[scale=0.4]{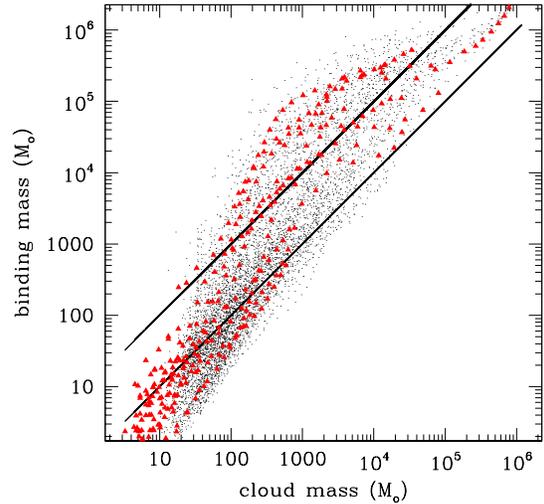}}
\caption{\label{bindmass}The relation between the binding mass needed to be gravitationally bound and the enclosed gas mass is plotted for various regions centred on local high densities throughout the Gravity re-simulation. The initial values are plotted as red triangles. Values determined for the same cloud but at increasing radii are plotted  individually. A  ratio of 1 (bound) and 10 (highly unbound) are plotted as solid lines. Only on small scales, and hence low masses, are the clouds bound even when self-gravity is included.}
\end{figure}

%In order to model star formation, we reran the previous simulation including the self-gravity of the gas to allow
 
In the {\sl Gravity} re-simulations, the inner 250 pc radius region of the Cloud re-simulations were reran near the maximum compression including self-gravity to follow the star formation process. 
 Figure~1 also shows three frames of the central regions from our standard $\Sigma_{\rm disc} \approx 4.2$ \solmas\ pc$^{-2}$  Gravity re-simulation, as seen from inside the plane of the galaxy.
%Figure~\ref{selfgrav1} shows the evolution, as seen in the plane of the galaxy,  of the self-gravitating simulation.
The clouds are  very structured with local star formation occurring along the filaments while large clusters form where the filaments intersect,
and are similar to star forming regions observed by the Herschel infrared space telescope \citep{Molinari2010,Hill2011}.  The clouds are not well defined objects with clear boundaries. Instead, the cold dense gas  represents the peaks in a continuous distribution.
Furthermore, gas enters and leaves these `clouds' throughout their lifetimes.

Star formation commences approximately $0.1$ Myr  after the start of the Gravity re-simulation, and continues for the full 5 Myr evolution. Local regions that  become self-gravitating have sizes of a few pc and thus densities $ > 100-1000\ \solm\ {\rm pc}^{-3}$.
These regions are colder in their centres with minimum temperatures reaching $<20$K.
The total star formation rate reaches a few times $10^{-2} \solm\ {\rm yr}^{-1}$.  The star formation efficiency, the fraction of the gas turned into stars, is a few per cent
in the first few million years and, in the absence of feedback and magnetic fields, increases to over ten per cent after five million years. 
%This value is high, at least partially due to our assumption of 100 \% efficiency within our sink particle radii of 0.25 pc. Star formation efficiencies
%within 0.25 pc are observationally found to be between 10 and 50 \%.

In addition to producing  the filamentary and clumpy structure in the gas, the shock drives large scale `turbulent' motions into the dense gas. The incoming
clumpy gas combined with the convergent flows in the spiral arm generate a large velocity dispersion ( see also \citealt{Bonnell2006,Heitsch2006, Koyama2002,Dobbs2007}). 
Figure~\ref{veldisp}
plots the velocity dispersion sizescale relation in the dense gas from near the beginning of the Gravity  re-simulation.  The velocity dispersion
is calculated as the mean of many different subregions of a given size centred on the densest gas particles. Indvidual subregions are calculated at increasing sizes until they meet another subregion or an edge in the mass distribution  is reached (when $M\propto R^n$ with $ n<0.75$).
These motions follow the standard
Larson-relations where the velocity dispersion scales as the square root of the size-scale of the region \citep{Larson1981,Heyer2004,Heyer2009}. In fact the
simulation reproduces both the approximate slope and normalisation ($\delta v= 0.9 R_{\rm pc}^{0.5}$ km/s) in the cold dense gas up to $\approx 10$ pc. 
Figure~\ref{veldisp} also shows the rms scatter in the data, plotted as errorbars showing that there is a significant variation in the local velocity dispersion.

The
velocity dispersion occurs as clumps entering the shock
on semi-random trajectories encounter differing amounts of gas and hence various levels of deceleration \citep{Dobbs2007}. Thermal instabilities
in the shocked gas also
aid in generating the observed velocity dispersions \citep{Heitsch2006, Koyama2002}. The velocity dispersion relation from the simulations becomes flat above $\approx 10$ pc, in contrast to that observed in GMCs. This could indicate that supernova or other feedback, or alternatively a magnetic coupling from the larger scale flows,  are needed to energise the ISM on larger scales.

The kinematic motions induced in the shock limit the onset of gravitational instabilities in the dense gas. 
Figure~\ref{bindmass} shows the relation between the cloud mass and the gravitational binding mass, the
mass needed  to be gravitationally bound given the internal kinematics.  We evaluate the cloud properties by integrating outwards from
the multiple dense regions and evaluate the total mass, kinetic energy and hence binding masses at each radius. The kinetic energies
are calculated in the centre of mass frame at each scale so that bulk motions are excluded. The resultant properties show that the clouds are bound only on small
sizescales and masses (masses below 100 to 1000 \solmas) and it is at these scales that we expect gravitational instabilities to induce star formation.
At large radii, the clouds tend to be unbound with binding masses of order a few to more than 10 times the cloud masses. This is broadly consistent with
observations of  Galactic molecular clouds \citep{Heyer2009}.  This excess kinetic energy at large scales can help maintain low star formation efficiencies
as significant parts of the clouds may never become gravitationally bound  \citep{Clark2008,Bonnell2011,Dobbs2011a}. 
The external pressure, including the ram pressure of the incoming gas
which is responsible for forming the clouds, is an important contributor on large scales. 

\begin{figure*}
%\centerline{\includegraphics[scale=0.4]{SFRslowreghigh925861.ps}
%\includegraphics[scale=0.4]{coldvstotalgassurdenearly.ps}}SFRfit861925
%\centerline{\includegraphics[scale=0.4]{SFR_comp_925_861.ps}
\centerline{\includegraphics[scale=0.4]{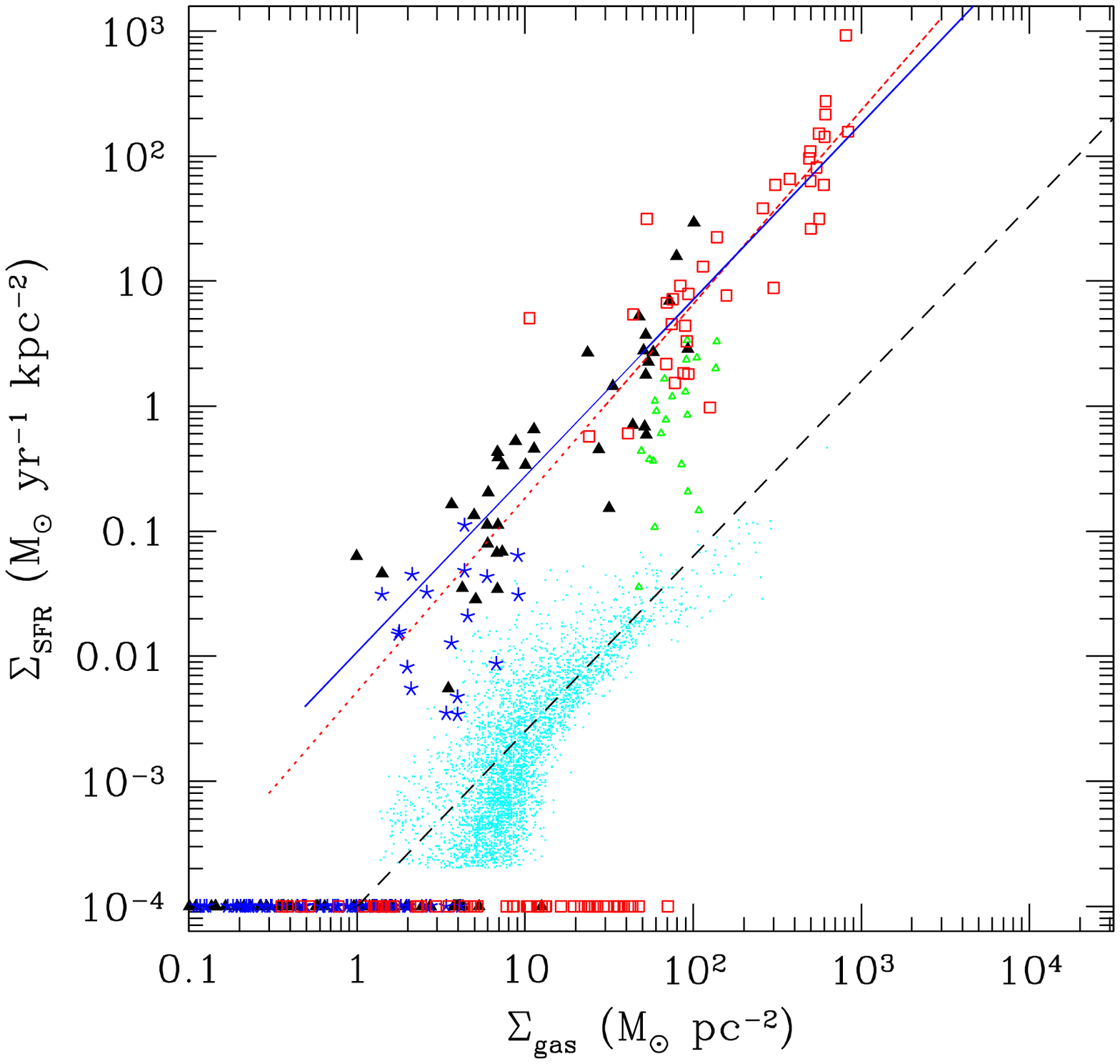}
\includegraphics[scale=0.4]{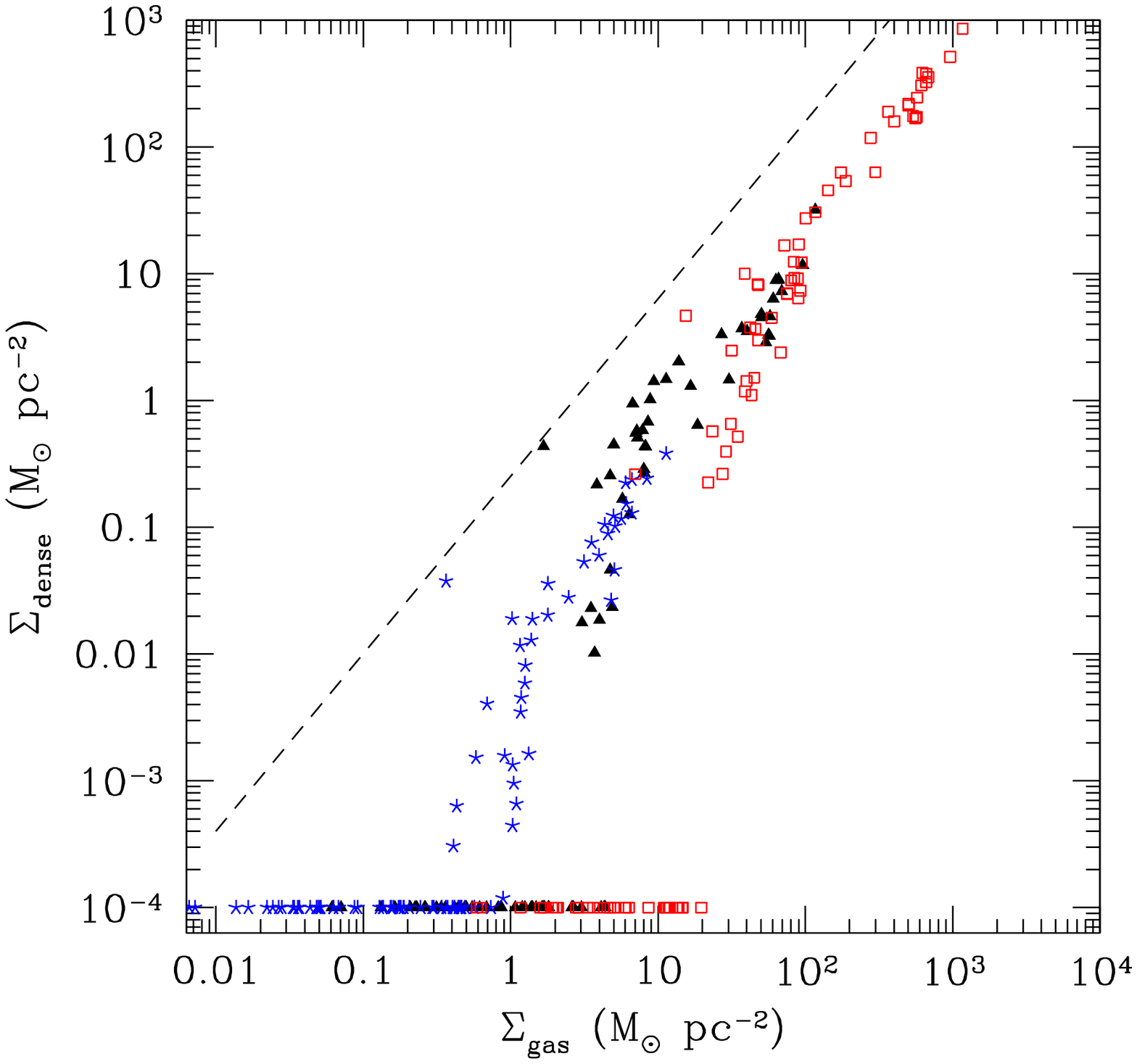}}
\caption{\label{SFRs}The local star formation rates per unit area  (left) and the surface density of cold dense gas (right) are plotted against the surface density of gas for the three {\sl Gravity}  re-simulations with initial surface densities of
0.42 \solmas\ pc$^{-2}$ (blue stars), $4.2 \solm\ {\rm pc}^{-2}$ (black filled triangles), and $42.0 \solm\ {\rm pc}^{-2}$ (red squares). The points represent local 50x50pc regions of the disc
as viewed from the plane of the galaxy. The SFRs  (left) correspond to a time
  $3.5 \times 10^6$ years from the start of the Gravity re-simulations, except for the high-surface density run which corresponds to   $1.3 \times 10^6$ years. The cold, $T<100K$, and dense, $\rho>10  \solm\ {\rm pc}^{-3}$, gas (right) is measured at the start of the Gravity re-simulations.
  The dashed line is the approximate location of  the Schmidt-Kennicutt $\Sigma_{\rm SFR} \propto \Sigma_{\rm gas}^{1.4}$ \citep{KennicuttEvans2012} relation (left) and shows   a $\Sigma_{\rm dense} \propto \Sigma_{\rm gas}^{1.4}$ relation (right). {The dotted and solid lines at the left show two $\chi$-squared fits to the data representing the total dataset (slope $= 1.55 \pm 0.06$) and for just the two higher surface density simulations (slope $=1.42\pm 0.07$)}. Null values  are set to $10^{-4}$ \solmas\ yr$^{-1}$ kpc$^{-2}$ and  \solmas\ pc$^{-2}$, respectively. 
The small cyan and green (small open triangles) points represent the \citealt{Bigiel2010} and \citealt{Heiderman2010} Table~1 data, respectively. Note that the  $\Sigma_{\rm dense} \propto \Sigma_{\rm gas}^{1.4}$  relation implies a molecular star formation relation of  $\Sigma_{\rm SFR} \propto \Sigma_{\rm dense}$.}
\end{figure*}

Although the clouds appear to be gravitationally unbound, there is significant collapse in the vertical direction throughout the {\sl Gravity} simulation. The turbulence
induced by the shock is anisotropic and is significantly lower in the vertical direction than in the plane of the galaxy. If supported only by the vertical motions, the dense gas would be gravitationally bound at all scales.
Feedback, magnetic fields and any hydrodynamical instabilities that may be missed by SPH could act to increase the vertical turbulence and hence decrease the star formation rates and efficiencies further.

\section{The Schmidt-Kennicutt relation}

One of the strongest observed relations of star forming galaxies is the Kennicut-Schmidt relation, where the surface density
of star formation is correlated with the surface density of gas as 
$\Sigma_{\rm SFR} \propto \Sigma_{\rm gas}^{1.4}$ \citep{Schmidt1959,Kennicutt1989,KennicuttEvans2012}.
In Figure~\ref{SFRs}, we plot the local rate of star formation (SFR), plotted in cells of 50 by 50 pc, against the total surface density of the gas in the {\sl Gravity}
re-simulation, as seen from within the plane of the galaxy. The star formation rates are calculated directly from the  mass incorporated into our sink particles over the timescale of the simulation.  

The star formation rates follow a Schmidt-Kennicutt  type $\Sigma_{\rm SFR} \propto \Sigma_{\rm gas}^{1.4}$ relation when
considering the total gas surface density  \citep{Schmidt1959,Kennicutt2007}. 
Included in Figure~\ref{SFRs} are the results for two other self-gravitating re-simulations, with initial surface densities a factor of ten higher (42 \solmas\ pc$^{-2}$) and lower  (0.42  \solmas\ pc$^{-2}$)
than the initial run, respectively.
Together there is a  strong continuous $\Sigma_{\rm SFR} \propto \Sigma_{\rm gas}^{n}$ type relation across all three {\sl Gravity} re-simulations,
arising solely due to the combination of the spiral shock, associated thermodynamics
and the self-gravity of localised regions. A best $\chi$-squared fit to the data finds $n \approx 1.55 \pm0.06$ when all non-zero values of the star formation rate are considered
over the three {\sl Gravity} simulations.  If we exclude the low surface density simulation, or fit only data with $\Sigma_{\rm gas} \ge 10$ \solmas\ pc$^{-2}$,  where a star formation  threshold may be present, we find a
best fit slopes of $n = 1.42 \pm 0.07$, and $n = 1.44 \pm 0.05$, respectively.

In contrast to the success in obtaining a reasonable slope for our star formation relation, the normalisation of the data is significantly higher than typical observed values.  
The star formation rates are offset to higher values from the Schmidt-Kennicutt relation, although they are closer to 
those found in nearby molecular clouds \citep{Evansetal2009,Heiderman2010} which also use direct estimates of the star formation rates. Averaging
over larger regions could reduce the relation if it included significant areas with gas but no star formation. This
is not the case in our present simulation as it was deigned to resolve the high-density star forming region.
It is likely that additional physics  is required to globally reduce the star formation rates.

\begin{figure}
%\centerline{\includegraphics[scale=0.4]{SFRslowreghigh925861.ps}
%\includegraphics[scale=0.4]{coldvstotalgassurdenearly.ps}}
\centerline{\includegraphics[scale=0.4]{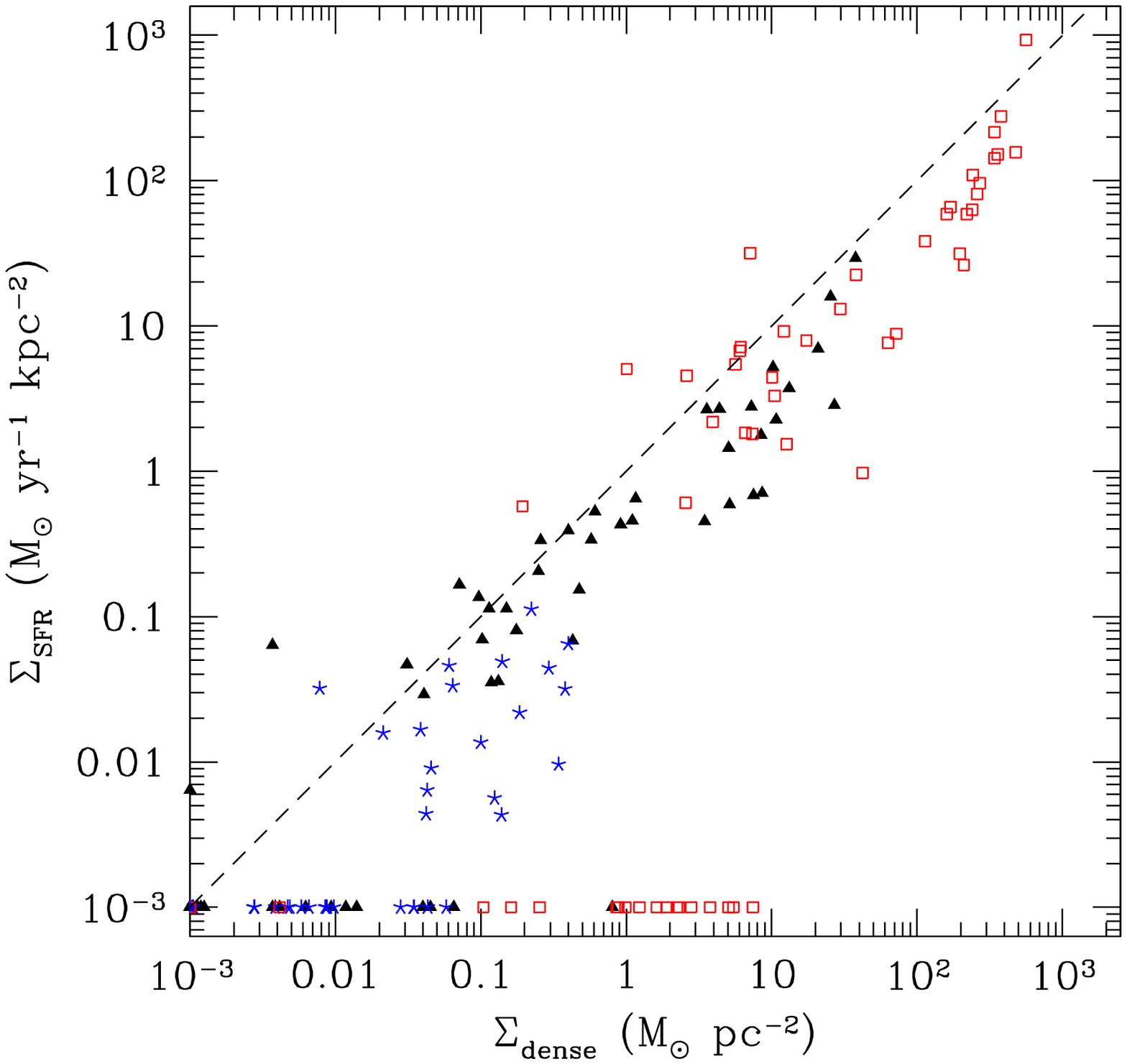}}
\caption{\label{SFRmol}The local star formation rates per unit area  are plotted against the surface density of cold ($T<100$) and dense ($\rho > 10  \solm\ {\rm pc}^{-3}$ gas for the three {\sl Gravity}  re-simulations with initial surface densities of
0.42 \solmas\ pc$^{-2}$ (blue stars), $4.2 \solm\ {\rm pc}^{-2}$ (black filled triangles), and $42.0 \solm\ {\rm pc}^{-2}$ (red squares). The points represent local 50x50pc regions of the disc
as viewed from the plane of the galaxy. The SFRs  correspond to a time
  $3.5 \times 10^6$ years from the start of the Gravity re-simulations, except for the high-surface density run which corresponds to   $1.3 \times 10^6$ years. The star formation rates 
  follow a quasi-linear relation with the surface density of dense gas (corresponding to molecular gas) consistent with  $\Sigma_{\rm SFR} = \Sigma_{\rm molecular}/t_{\rm ff}$ and 
  a free-fall time of 1  to 10 million years. These simulations do not include feedback or magnetic fields which would slow the star formation process.}
\end{figure}

Many surveys have shown that the star formation law is shallower when looking at molecular gas with a direct $\Sigma_{\rm SFR} \propto \Sigma_{\rm molecular}$ relation \citep{Bigiel2008,Bigiel2011,Schruba2011}.
This suggests that the  star formation rates are linked to the formation of dense molecular gas and that thereafter there is a direct correspondence  at a
constant rate given by the free-fall time and an efficiency factor  $\Sigma_{\rm SFR} = \epsilon \Sigma_{\rm molecular}/t_{\rm ff}$ (e.g. \citealt{Krumholz2009}).
We also find (see Figure~\ref{SFRmol}  a near-linear relation of the star formation rates to the cold   ($T<100$K) and dense ($\rho>10  \solm\ {\rm pc}^{-3}$)  gas  $ \Sigma_{\rm dense}$. This occurs as the free-fall time of the cold and dense gas is roughly constant at a few to 10 Myrs.  We expect that feedback and magnetic fields
or an increase in the vertical component of turbulence should increase  this timescale for star formation and decrease the star formation rates in a global fashion. 

If the star formation rates follow a linear relation with the dense, cold,  molecular gas, then it is at the phase where molecular clouds are formed that the non-linear nature of the Schmidt-Kennicutt relation must arise.
In Figure~\ref{SFRs}, we also plot  the surface density of cold ($T<100$K) and dense ($\rho>10  \solm\ {\rm pc}^{-3}$)  gas  $ \Sigma_{\rm dense}$, that corresponds to molecular clouds,
as a function of the total surface density of gas as viewed horizontally in the plane of the galaxy.  We find a $\Sigma_{\rm dense} \propto \Sigma_{\rm gas}^{1.4}$ relation
with a similar slope as the star formation Schmidt-Kennicutt relation. This shows that the non-linear slope in our simulated S-K relation arises in the formation of the cold gas.
The `molecular' gas fraction  is zero below surface densities of a few \solmas pc$^{-2}$ and increases linearly with the surface density from a few percent to being predominantly cold and dense at high surface densities (see also \citealt{Krumholz2009}). 

Unexpectedly, the $\Sigma_{\rm dense} \propto \Sigma_{\rm gas}^{1.4}$ relation is found in both the self-gravitating and the non self-gravitating simulations. 
That this fundamental relation arises independently of self-gravity shows that gravity cannot be the determining physics in generating the cold gas,
and ultimately in setting the star formation rates.
Instead,  it is the shock and thermal physics of how the
cold dense gas is generated which drives the resulting star formation rates (see also \citealt{GloverClark2012a,GloverClark2012b,Krumholz2011,Krumholz2012}).
Gravity's role is then limited to the collapse of the dense gas (which in our simulations has a near constant free-fall time, $t_{\rm ff}$) such that 
$\Sigma_{\rm SFR} \approx \Sigma_{\rm dense}/ t_{\rm ff} \propto \Sigma_{\rm dense}$. A slower collapse due to magnetic fields or equivalently a relaxation of our assumption of 100 per cent star formation efficiency
would then explain the offset found in the star formation rates.

\begin{figure}
\begin{center}
\includegraphics[width=8cm]{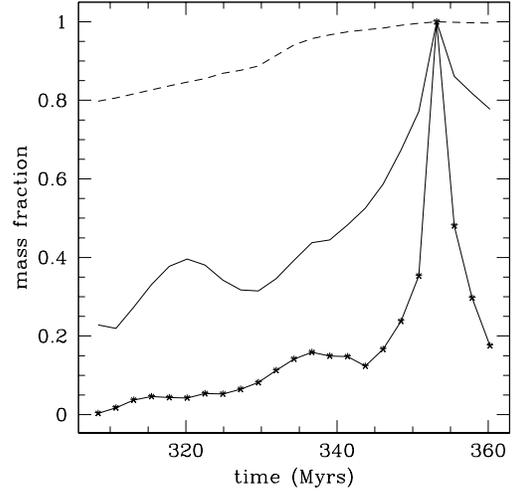}
\caption{\label{massfractioncold} The mass fraction of cool ($T<1000$ K,  dashed line), cold ($T<100$ K, solid line) and dense gas ($\rho > 10 \solm {\rm pc}^{-3}$) gas is plotted for a region of high star formation throughout the {\sl Cloud} re-simulation. The gas spends only a short time (~5 Myr) as cold and dense and hence capable of undergoing gravitational collapse and forming stars. In contrast, the gas is mostly cool ($T<1000$ K) throughout the evolution before the spiral shock compresses the cool gas to form the dense cold gas where star formation occurs (see also \citealt{Dobbs2012}).}
\end{center}
\end{figure}

The steeper relation between the dense gas and the total gas present must arise  from the shocks and accompanying cooling. In the re-simulations it is found that the cold dense gas is formed by collisions between clumps of cool ($100 <T< 1000$ K) gas formed in previous shocks.  Figure~\ref{massfractioncold} shows the evolution of gas that comprises a region of high star formation rate over the 50 Myr
of the {\sl Cloud} re-simulation. The gas spends only a relatively short time (5 Myr) as  dense and cold ($T<100K$; $\rho>10  \solm\ {\rm pc}^{-3}$)  but is mostly cool ($T<1000$ K) throughout the full 54 Myr of the re-simulation.
This `cool' gas (often termed the warm neutral medium WNM) is due to earlier shocks during the spiral arm passages and forms a clumpy medium embedded in the warmer gas (see also \citealt{Dobbs2012}). Subsequent collisions involving this cool gas result in much stronger compressions, hence the formation of the cold, dense ($T<100K$; $\rho>10  \solm\ {\rm pc}^{-3}$) gas and ensuing star formation. Where the probability
of an individual  clump colliding with another is low, this results in a non-linear scaling between the incoming $\Sigma_{\rm gas}$ and shock-produced $\Sigma_{\rm dense}$.
We construct below a  simple clumpy shock model based on the requirement of clumps to interact in order to form the dense gas. This toy model can reasonably reproduce the $\Sigma_{\rm dense} \propto \Sigma_{\rm gas}^{1.4}$ relation when clumps are in pressure equilibrium with the gas and their sizes are then given by $R\propto \Sigma_{\rm gas}^{-1/3}$.

\subsection{\bf Star formation rates: thresholds}

Both the star formation rates and the surface density of cold gas (Fig. \ref{SFRs}) exhibit minimum surface densities below which there is no dense gas or star formation. These thresholds
are due to a minimum external pressure  necessary to maintain the cool (WNM) gas, from which both the dense gas and the star formation activity are derived. At surface densities lower than $\Sigma_{\rm gas}\approx 1 \solm$ pc$^{-2}$, the dense gas is over-pressured relative to the external medium and expands. It
then heats up as the decreasing cooling rates at lower densities no longer balance the external radiation field.  
This threshold in maintaining the cool gas produces a steep
$\Sigma_{\rm SFR} - \Sigma_{\rm gas}$ over a short range. 
The higher threshold in the $42 \solm$ pc$^{-2}$  resimulation reflects simply a scaled up minimum surface density where cool gas is present in the initial conditions.

The thresholds occur at higher  surface densities in the re-simulation that started with a higher mean surface density. This
is likely to be due to the resolution and the methodology used to construct the re-simulations. The initial amount of cool gas
present in each re-simulation was taken from the global disc model and then scaled accordingly. For the lower surface density
re-simulation, this will have produced an overabundance of cool gas but this will quickly dissipate
as it will no longer be in pressure equilibrium with the lower external pressure. In contrast, the higher surface density re-simulation
will start with a deficit of cool gas. Given a sufficiently long evolution, this cool gas would be self-consistently regenerated
but it is apparent that over the 50 Myrs of the re-simulations, this has not occurred. Thus, there is a physical threshold that emerges
from the simulations but in the high surface density re-simulation has been scaled  to higher surface densities due to the resolution.

\begin{figure}
\begin{center}
\includegraphics[width=8cm]{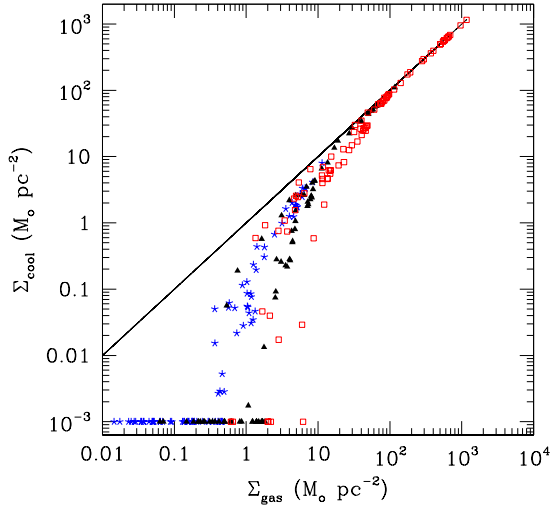}
\caption{\label{coolgas} The surface density of cool gas ($T<1000$ K) relative to the total gas density for the Cloud Re-simulations with mean surface densities 0.42 \solmas\ pc$^{-2}$ (blue stars), $4.2 \solm\ {\rm pc}^{-2}$ (black filled triangles), and $42.0 \solm\ {\rm pc}^{-2}$ (red squares). Above the threshold needed to sustain cool gas the surface density of cool gas scales almost linearly with the total surface density.}
\end{center}
\end{figure}

\section{\bf Star formation rates: a clumpy shock model}

The star formation rates scale directly with the surface density of cold gas. We therefore need to understand the relation between the dense gas and the total surface density $\Sigma_{\rm dense} \propto \Sigma_{\rm gas}^{1.4}$ . As this occurs even in the absence of self-gravity, we require a model that includes only the shock compression and cooling. In our simulations the cold gas is primarily produced by the collision of cool clumps of gas which are compressed to higher density cold gas by the additional ram pressure. The likelihood of cold gas formation is therefore determined by the probability of cool gas collision.

The amount of cold, dense gas generated in a shock $\Sigma_{\rm dense}$ is given by the amount of cool gas entering the shock and the probability of an individual clump of
cool gas hitting another clump of cool gas, so
\begin{equation}
\Sigma_{\rm dense}\propto\Sigma_{\rm cl} P
\end{equation}
where $\Sigma_{\rm cl}$ is the surface density of cool clumps and $P$ is the probability of each clump colliding with another. Let us consider the maximal and minimal cases which the true collision probability, $P$, must lie between. In the regime where the clumps occupy a large fraction of the total volume, all clumps will hit another one, $P\approx 1$, and the rate of forming dense gas from these collisions will scale as $\Sigma_{\rm cl}$. In the regime where the clump volume fraction is low, the probability of a given clump colliding with another is then $\Sigma_{\rm cl} \pi R^2$, where $\pi R^2$ is the cross section for collision of a clump of radius $R$. If $R$ is independent of environment, then the rate of forming dense gas from the clumpy shock will scale as $\Sigma_{\rm cl}^2$. As the true clump collision probability lies between these two regimes, the formation of dense gas must scale as the cool gas surface density to a non-linear power between 1 and 2.

To estimate the value of the true collisional rate we consider the following simplified toy model. Firstly we assume that the cool gas clumps are in pressure equilibrium with their environment, and that all the clumps have a similar gas temperature. This means that the radius of a given clump, R, will be proportional to its surrounding pressure, p, which is proportional to the gas surface density as $R\propto p^{-1/3} \propto\Sigma_{gas}^{-1/3}$. Putting this into our above relations this gives
\begin{equation}
\Sigma_{\rm dense}\propto\Sigma_{\rm cl}^2 \Sigma_{gas}^{-2/3}.
\end{equation}

Figure \ref{coolgas} shows the surface density of cool clumpy gas verses the total gas density. As previously discussed, below a threshold density of around 1 M$_\odot$pc$^{-2}$ there is insufficient pressure to confine the cool gas, however above this threshold the surface density of cool clumpy gas scales close to linearly with the gas density, $\Sigma_{cl}\propto\Sigma_{gas}$. The dense gas surface density then becomes roughly
\begin{equation}
\Sigma_{\rm dense}\propto \Sigma_{gas}^{4/3}
\end{equation}
in close agreement with the relationship found in our full simulation.

\begin{figure}
\begin{center}
\includegraphics[scale=0.4]{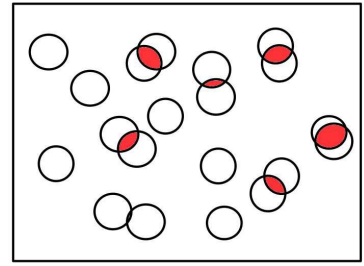}
\caption{\label{overlap} To test the probability of cold gas generation from collisions we randomly place circles of radius $R\propto\Sigma_{gas}^{-n}$ in a given area and calculate the surface density of dense gas where they overlap.}
\end{center}
\end{figure}

\begin{figure}
\begin{center}
\includegraphics[scale=0.4]{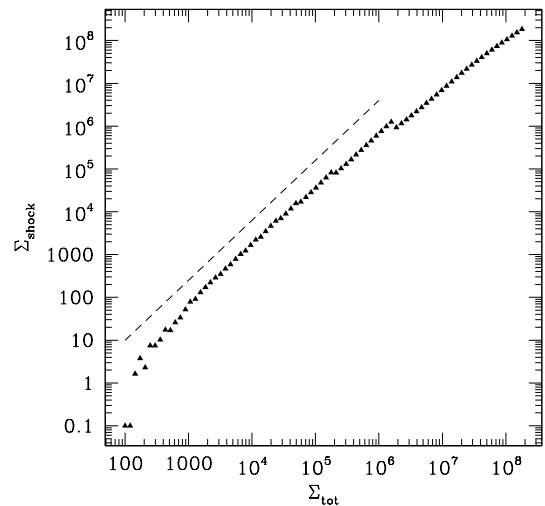}
\caption{\label{shockmodel} The resultant surface densities (in dimensionless units) of dense (shocked) gas versus total gas for the simple clumpy shock model where $R\propto \Sigma^{-1/3}$.
A clumpy shock model can reproduce a reasonable relation between the surface density of dense gas and  total surface density provide
$R\propto \Sigma^{-n}$ with $n$ in the range of  $1/3 {\rm\ and\ } 0.4$. The dashed line shows a slope of $1.4$.}
\end{center}
\end{figure}

To further investigate these ideas we make the following test of the probability of randomly generated clumps overlapping. In a given area we randomly put down clumps with a radius $R\propto\Sigma_{gas}^{-n}$. Where clumps overlap we assume cold dense gas will be produced and compare its surface density with the total cool gas surface density. Figure \ref{overlap} shows a schematic of this process. 
%We repeat the test with values of $n$ between $2/9$ and $0.5$ and find that \Sigma_{\rm dense}\propto\Sigma_{gas}^{1.4} when $n$ is between $1/3$ and $0.4$. Figure \ref{shockmodel} shows the relationship for $n=1/3$. 
We repeat the test varying $n$ and find reasonable $\Sigma_{\rm dense}\propto\Sigma_{gas}^{1.4}$ relations when $n$ is between $1/3$ and $0.4$. Figure \ref{shockmodel} shows the relationship for $n=1/3$.

\section{Conclusions}

The results presented here show that a spiral shock coupled with the cooling rates in the interstellar medium can explain the origin of the Schmidt-Kennicutt relation for star formation rates in galaxies. The shock formation of cold gas
provides the link between the total gas surface density and the ensuing star formation rates. It also explains the linear relation found between the dense molecular gas and the star formation rates, such that $\Sigma_{\rm SFR}\propto \Sigma_{\rm dense}\propto \Sigma_{\rm gas}^{1.4}$. 

Self-gravity is only required to operate locally in the dense gas which has a characteristic free-fall time. 
The presence of cooler gas ($T<1000$K)
in pressure equilibrium with the warm gas as it enters the shock is important in forming the dense, cold  self-gravitating clouds where star formation occurs \citep{Pringle2001}. A minimum pressure or surface density
is required to maintain the cool gas, and hence sets a threshold for star formation.   A simple shock model for the pressure-bounded cool gas can account for the non-linear nature of the Schmidt-Kennicutt relation.

These results are  shown here for the case of a grand-design type spiral galaxy but the nature of the shock-driven cooling should apply to any
galactic driven flow. We expect similar processes to occur in flocculent spirals  or other convergent flows, and hence provide a universal explanation for the triggering of star formation and the ensuing rates.
The necessary condition is of a bi-stable interstellar medium where cooling instabilities occur above a critical density. 

The actual values of the star formation rates found here are offset to higher values from the extragalactic values.  Additional forms of support including higher vertical turbulence, magnetic fields and feedback
could easily lower the star formation rates and hence account for the discrepancy, as would the sudden turning on of self-gravity. Magnetic
fields and feedback will delay the onset of star formation and 
decrease the subsequent rates \citep{Banerjee2009,Price2009,Vazquez2010}, while the sudden turning on of self-gravity neglects any previous star formation histories in the gas
and thus artificially increases the rates. Taken together, these effects will 
necessitate higher gas surface densities to produce the same star formation rates, but given the success here of reproducing the  $\Sigma_{\rm SFR} \propto \Sigma_{\rm gas}^{1.4}$, they need only provide  a constant
offset.

%\section*{Acknowledgments}

\section*{Acknowledgements} 
We thank Lee Hartmann, Neal Evans, Frank Bigiel and Mark Heyer and the referee for comments that improved the text.
IAB thanks the Laboratoire d'astrophysique de Bordeaux for hosting part of this research.
IAB acknowledges funding from the European Research Council for the FP7 ERC advanced grant project ECOGAL.
CLD acknowledges funding from the European Research Council for the FP7 ERC starting grant project LOCALSTAR.
RJS acknowledges support for grant SM321/1-1 from the DFG Priority Program 1573, ``The Physics of the Interstellar Medium''.

%\bibliography{/Users/mbate/Documents/home2/Tex/Papers/Bonnelletal2010/mbate}

\bibliography{iab}
%\bibliography{scibib}

\bsp

\label{lastpage}

\end{document}